# Getting through the nature of silicene: $sp^2$-$sp^3$ two-dimensional silicon nanosheet

*Eugenio Cinquanta*[*,1], *Emilio Scalise*[2], *Daniele Chiappe*[1], *Carlo Grazianetti*[1,3], *Bas van den Broek*[2], *Michel Houssa*[2], *Marco Fanciulli*[1,3] *and Alessandro Molle*[1]

[*] Dr. E. C. Corresponding-Author: email: eugenio.cinquanta@mdm.imm.cnr.it

[1] Laboratorio MDM, IMM-CNR, via C. Olivetti 2, I-20864 Agrate Brianza (MB), Italy

[2] Semiconductor Physics Laboratory, Department of Physics and Astronomy, University of Leuven, B-3001 Leuven, Belgium

[3] Dipartimento di Scienza dei Materiali, Università degli Studi di Milano Bicocca, via R. Cozzi 53, I-20126, Milano (MI), Italy

Abstract

By combining experimental techniques with ab-initio density functional theory calculations, we describe the Si/Ag(111) two-dimensional system in terms of a $sp^2$-$sp^3$ crystalline form of silicon characterized by a vertically distorted honeycomb lattice. We show that 2D $sp^2$-$sp^3$ Si NSs are qualified by a prevailing Raman peak which can be assigned to a graphene-like $E_{2g}$ vibrational mode and that highly distorted superstructures are semiconductive whereas low distorted ones behave as semimetals.

In the latest years a considerable effort has been devoted to identify the graphene-like allotrope of silicon, namely silicene, both from the experimental and theoretical point of view.[1-5] Although free-standing (FS) silicene has been hypothesized as a pure $sp^2$ hybridization of silicon, its existence in ambient conditions has been argued due to strong electron correlation.[6] On the other hand, many authors succeeded in epitaxially growing a two-dimensional (2D) Si nanosheet (NS) on the Ag(111),[1,7-9] Ir(111)[10] and $ZrB_2$[11] surface with differently buckled $sp^2$-$sp^3$ arrangements.

Compelling evidences of Dirac fermions and preliminary indications of band gap opening in a Si NS epitaxially grown on Ag(111) substrates[1] makes it enormously attractive in terms of fundamental properties[2-5] and technological transfer.[2,11]

So far epitaxial silicon NSs have been referred to as silicene layers because their topography is consistent with a distorted, energetically stable, honeycomb lattice.[1,8] However, for a proper identification of these Si NSs as graphene-like 2D materials, it is essential to understand the arrangement of the bonds in the Si NSs (i.e. geometry and hybridization) and how the Si bonds influence their vibrational and electronic properties.[1,12]

To get through this aspect and elucidate the intimate nature of 2D Si NSs, various superstructures[7,8,13-15] of a Si monolayer epitaxially grown on Ag(111) substrates have been scrutinized and selected by means of in situ scanning tunneling microscopy (STM) and, after non-reactive encapsulation,[16] their vibrational modes have been unveiled by Raman spectroscopy. Experimental results have been then interpreted with *ab-initio* Density-Functional Theory (DFT) calculation.

Epitaxial 2D Si NSs on Ag(111) evidence a multi-phase character,[7-9,13-14] namely a variety of reconstructed Si domains with characteristic periodic patterns including 4x4, √13X√13-II, and 2√3X2√3-II (where the periodicity is defined with respect to the Ag unit cell) which can be recognized by STM[7-9] or electron diffraction techniques.[1,13-14] Each superstructure corresponds to a honeycomb Si domain with characteristic buckling γ, i.e. the local vertical displacement induced by the underlying Ag atoms,[14, 15] and follows a well-defined phase diagram dictated by the deposition temperature for a fixed deposition flux.[8, 13, 14, 17] In our experimental approach two characteristic growth regimes were identified: one at a lower deposition temperature (T=250°C) where the 4x4 and the √13X√13-II Si superstructures coexist on the Ag(111) surface, and a second one at a higher temperature (T=270°C) where a uniform Si layer with a characteristic 2√3X2√3-II superstructure is observed.

The topography of the Si NS grown at 250°C is shown in Fig. 1(a) therein evidencing 2D Si domains with lateral size in between 20nm and 50nm and uniformly distributed throughout the Ag(111) surface. The atomically resolved STM magnification in Fig. 1(b) shows that Si atoms are sequentially placed either in between or on top of the Ag atoms, thus generating two characteristic surface patterns, a 4x4 (green contour in Fig. 1.b ) or a √13X√13-II buckled superstructure (red contour in Fig. 1.b).[14, 15]

The 4x4 superstructure is defined with respect to the unit cell of the underlying Ag(111) surface. The calculated cell parameter of the 4x4 superstructure is 11.78 Å, i.e. nearly four times the cell parameter of the Ag(111) surface. The 4x4 supercell includes three hexagonal unit cells of silicene with cell parameter of 3.926 Å. This value is larger by about 1.5% than the calculated value for the FS silicene[3] thus indicating that the epitaxial silicene on Ag is slightly stretched. Other relevant features discriminating the epitaxial silicene from the FS one are: a) the more pronounced vertical stacking ($0.71<γ<0.79$ Å), and b) the non-uniform distribution of buckled atoms. The latter feature is a consequence of the substrate-induced breaking of the threefold rotational symmetry present in the FS silicene. In detail, only 6 of the 18 atoms building up the 4x4 superstructure are on a higher plane with respect to the lower one which is closer (~2 Å) to the Ag substrate and contains the remaining 12 atoms.[1, 7, 8, 13-15]

The √13X√13-II reconstruction is a supercell of silicene made up of 14 atoms and rotated by about 5.2° relatively from its initial position in the 4x4 superstructure. The angle between the axis of the √13X√13-II superstructure and the 4x4 one is 13.9°. Only 4 of 14 atoms are on the higher plane, forming a buckled structure with a vertical parameter very close to that of the 4x4 superstructure ($\gamma$~0.79 Å).[7, 8, 13-15]

Figure 1.d reports the large scale topography of the 2√3X2√3-II Si NS superstructure grown at 270°C. For coverage $\vartheta$~1 ML the 2√3X2√3-II domains extend over the whole Ag(111) terraces thus proving an improved continuity of the layer with respect to the more fragmented surface structure of the Si NS in Fig. 1(a). The atomic scale topography in Figure 1.e exhibits the characteristic surface patterns of the 2√3X2√3-II superstructure, as reported elsewhere.[7, 8, 14, 15] This superstructure is characterized by a unit cell with 12 planar and two buckled ($\gamma$~1 Å) atoms and it is misaligned from the 4x4 superstructure by an angle of 30°.[14, 15]

Remarkably, for the three considered superstructures the bond length ranges from 2.34 to 2.39 Å for the 4x4 one, from 2.31 to 2.36 Å for the √13x√13-II one and from 2.28 and 2.37 for the 2√3x2√3-II, in all cases being quite close to the Si-Si bond length in $sp^3$ diamond like Si (2.34 Å). DFT modeling of these superstructures generally results in a bond angle distribution peaked around 109° and 120° which respectively identify $sp^3$ and $sp^2$ hybridized Si atoms. It must be emphasized that the 2D Si NSs preserve the $sp^2$ trigonal geometry (each Si atoms is coordinated with other 3 Si atoms) whereas the bond lengths and bond angles evidence a $sp^3$-like nature because of the presence of buckled Si atoms stemming from substrate-induced local vertical distortions. More specifically, the substrate constraints concomitantly induce i) a non-uniformly distributed vertical alignment of Si atoms, ii) a horizontal "in-plane" strain being tensile in character, and iii) a partial hybridization of the resulting unhybridized Si $P_z$ orbitals with the underlying Ag atoms.[12] These features are expected to dramatically impact the vibrational properties of each single superstructure and they can be elucidated by means of Raman spectroscopy.

Figure 2.a shows the Raman spectrum of the sample including the 4x4 and the √13X√13-II superstructures.

The spectrum is characterized by an intense peak located at 516 cm$^{-1}$ presenting an asymmetric and broad shoulder at lower frequency (440-500 cm$^{-1}$). It is interesting to notice that this feature cannot be provided by a full sp$^3$ nanocrystalline silicon (nc-Si), because only ~2nm tailed nanocrystals would induce the observed redshift,[18, 19] whereas the STM probed domains turn out to have a width ranging from 20 to 50 nm (see Figure 1.a). A derivation of the peak at 516 cm$^{-1}$ from a fully sp$^3$ stressed Si can be also ruled out because the periodicity of the Si lattice (0.54 nm) is by far larger than the Ag(111) surface cell parameter (0.29 nm). As a consequence, the resulting compressive strain should up-shift Raman peak of bulk silicon at 520 cm$^{-1}$, thus not being consistent with the observed peak position.[20] The two extra-features present at ~300 and ~800 cm$^{-1}$ are provided by the $Al_2O_3$ capping layer.[21]

In order to interpret the experimental spectrum in Fig. 2.a, the non-resonant Raman spectrum of the defect-free 2D Si NSs on Ag(111) has been simulated and reported in Fig. 2b and 2c. In detail, Fig. 2.b and 2.c show the vibrational modes of both √13X√13-II and 4x4 superstructures, respectively. Despite the sp$^2$-sp$^3$ nature of the two phases, a doubly degenerate $E_{2g}$ mode is predicted to take place at 505 cm$^{-1}$ and 495 cm$^{-1}$ for the √13X√13-II and 4x4 superstructures respectively. These frequencies are in pretty good agreement with the experimental feature at 516 cm$^{-1}$, taking into account an underestimation of about 1-5% for the vibrational frequencies, typical for the implemented computation (see methods for details).[22, 23] As for the graphene G-peak, this mode is provided by the bond stretching of all sp$^2$ silicon atom pair, thus reflecting the 2D-Si NSs sp$^2$ character and being the fingerprint of a honeycomb lattice. On the other hand $E_{2g}$ mode frequency, 516 cm$^{-1}$, is strictly dependent on the sp$^3$-like Si-Si bond length (~2.34 Å), being present also in the FS silicene but with relatively much higher frequencies (~570 cm$^{-1}$) as due to the shorter and full sp$^2$ bond length (2.28 Å).[23] Although the presence of $E_{2g}$ modes clearly indicates the honeycomb nature of the 2D Si-NSs lattice, the broad and asymmetric shoulder in the 440-500 cm$^{-1}$ is not compatible with a defect-free planar trigonal geometry. Hence it is interesting to elucidate the presence of this feature, especially considering the structural complexities characterizing 2D Si-NSs, as previously discussed (Fig. 1).

It is useful noticing that defect-free graphene, characterized by a planar honeycomb lattice, provides a Raman spectrum characterized by the presence of a $E_{2g}$ mode (G peak at 1581 cm$^{-1}$) plus the 2D peak at 2600 cm$^{-1}$ activated by inter-valley electron-phonon scattering between K and K' point of the first Brillouin Zone (FBZ).[25] For graphene, the presence of an additional D($A_{1g}$) peak at ~1300 cm$^{-1}$ is expected when double resonance processes take place, that is intra-valley electron-defect scattering at K in the FBZ.[24]

Surprisingly, unlike FS silicene,[23] the calculated spectra of the Si superstructures are characterized by several vibrational modes having a non-vanishing Raman intensity (denoted as D, T and K in Fig. 2.b and 2.c) also assuming a defect-free configuration in the model.

Interestingly, for the 4x4 case, two $A_{1g}$ modes at 436 cm$^{-1}$ and at 466 cm$^{-1}$ (reported as D and T respectively in Fig. 2.c) are Raman-active and, surprisingly, the D one is the dominant feature.

Similarly, the √13X√13-II superstructure presents, beyond the $E_{2g}$ vibrational mode, a D ($A_{1g}$) peak at 455 cm$^{-1}$ plus a K ($B_{2u}$) at 475 cm$^{-1}$, as reported in Fig. 2.c.

By comparing the calculated spectra reported in Fig. 2.b and 2.c, it can be noticed that the $A_{1g}$ mode at 436 cm$^{-1}$ is the most intense Raman-active mode in case of the 4x4 superstructure, while the $E_{2g}$ mode becomes largely dominant for the √13X√13-II one. Since these two superstructures mutually differ in terms of their intimate buckling distribution, the observed intensity variation of the Raman features is reasonably related to their different atomic configuration. In particular, a reduced number of buckled bonds (as in the √13X√13-II) results in a strong suppression of the $A_{1g}$ modes intensity with respect to the $E_{2g}$ peak. The origin of $A_{1g}$ mode activation can be then associated to an intrinsic "disorder" related to the non-uniform substrate-induced buckling and to the mixed sp$^2$-sp$^3$ nature of the honeycomb Si lattice. More in details, for the 4x4 case, the D($A_{1g}$) mode comes form the breathing-like displacement of planar hexagons, while the T($A_{1g}$) is related to the breathing-like displacement of non-planar hexagons (see Fig. S1 in Suppl. Information); for the √13X√13-II, the D($A_{1g}$) and K($B_{2u}$) modes arise from the breathing mode of the hexagonal rings having alternating up and down-standing atoms and to Kekule-distorted[25] hexagonal rings respectively (see Fig. S1 in Suppl. Information).

Since the experimental Raman spectrum integrates the contribute of the two phases weighted by their abundances ratio, the Raman spectrum of 2D sp$^2$-sp$^3$ Si NS is dominated by the $E_{2g}$ modes of the two superstructures, along with the asymmetric shoulder provided by the interplay of the disorder-activated modes ($A_{1g}$ and $B_{2u}$).

Remarkably, the experimental spectrum in Fig. 2a is well reproduced by the calculated spectrum of the √13X√13-II superstructure, rather than by the one of the 4x4 one. This asymmetry may come from the smaller amount of 4x4 superstructure domains with respect to the

√13X√13-II superstructure ones. Nonetheless, it is interesting to explore whether a Raman resonance, not implemented in the adopted DFT framework, might take place also in consideration of the measured band gap of 0.6 eV for the 4x4 superstructure. In this case, resonance effects are expected to selectively amplify the $E_{2g}$ and $A_{1g}$ Raman-active modes. Indeed, the former is an intra-valley phonon scattering process at Γ and thus its resonance is strictly related to electronic transitions when the incident radiation is properly tuned with the direct band gap transition, with the consequent enhancement of the related Raman signal.

To get through possible resonant behaviors, the Raman spectrum of the two configurations has been investigated as a function of the excitation energy. The case of the mixed 4x4/√13X√13-II is reported in Figure 3. The pronounced enhancement of $E_{2g}$ intensity with increasing excitation energy indicates a semiconductive character (formally similar to the one of $sp^3$ silicon),[26] being consistent with the reported presence of a direct gap in the 4x4 superstructure.[1] Interestingly, an additional hint at the multiphase character of the sample comes from the frequency dispersion of the band at ~900 cm$^{-1}$: the observed blueshift as a function of the excitation energy is similar to the one observed for the 2D peak of graphene, which is provided by the presence of the Kohn anomaly in the K point of the FBZ.[24] Second order Raman feature of bulk-Si also places in between 900 and 1000 cm$^{-1}$ as due to two transverse optical modes. When varying the exciting frequency, this band is affected by a change of the peaks intensity ratio because of Raman resonance[26] rather than by frequency dispersion.

The resonance behavior and the frequency dispersion can be rationalized by attributing a semiconducting character to the 4x4 superstructure consistently with ARPES outcomes[1] and a graphene-like character to the √13X√13-II one. However, it is not trivial to discriminate the two contributions as, even finely tuning the deposition parameters, it was not possible to isolate just one of these two superstructures. In this effort, the 2√3X2√3-II superstructure has been successfully stabilized as single phase after carefully tailoring the growth conditions (see Fig. 1.d). Figure 4 compares the experimental Raman spectrum of the 2√3X2√3-II (Fig. 4.a), acquired with four excitation energies, with the calculated one (Fig. 4.b). As for the √13X√13-II superstructure, the Raman spectrum is characterized by the presence of a strong $E_{2g}$ mode at 521 cm$^{-1}$, confirming both the sp$^2$ hexagonal lattice symmetry and the sp$^3$ like nature of its bonds lengths. Furthermore weak $A_{1g}$ modes are also active (see Fig. S2 in the Suppl. Information), thus indicating a lower amount of intrinsic-disorder with respect to the previously analyzed superstructures. Interestingly the Ultra-Violet (UV) Raman spectrum exhibits a quite low signal/noise ratio, thus reflecting a very low cross section at this frequency, that is opposed to the expected behavior of the sp$^3$ diamond-like silicon.[26] By observing the spectra reported in Fig. 4a, two remarkable facts must be underlined: a) no Raman resonance affects the $E_{2g}$ mode, and b) a frequency dispersion characterizes the ~900 cm$^{-1}$ band. Both observations are not compatible with a full semiconductive sp$^3$ structure,[26] but they rather reflect the presence of a Si honeycomb lattice with a characteristic graphene-like behavior, that is a non-resonant behavior and Kohn-anomaly related frequency dispersion.

In summary, by combining experimental techniques with ab-initio DFT calculation, we describe the Si/Ag(111) 2D systems in terms of a sp$^2$-sp$^3$ form of silicon characterized by a vertically distorted honeycomb lattice provided by the constraint imposed by the substrate. The Raman spectrum reflects the multi-hybridized nature of the 2D Si NSs resulting from a buckling-induced distortion of a purely sp$^2$ hybridized structure. This sp$^2$-sp$^3$ character provokes an intrinsic disorder which leads to the activation of Raman-inactive vibrational modes ($A_{1g}$), whose intensity is a function of the amount of buckled Si atoms. For the 2D Si-NSs where two different superstructures coexist (4x4 and √13X√13-II), the dependence of the Raman response with the excitation energy allows us to recognize a non-trivial semiconducting character. We then successfully isolate the 2√3X2√3-II superstructure which makes evidence of a low-distorted honeycomb lattice thus opening new interest for 2D $sp^2$-$sp^3$ silicon nanosheets with a graphene-like symmetry and electronic character.

**Methods**

The synthesis of our sample is the same adopted in[7] and briefly reported in the Supplementary Information. In order to prevent silicon oxidation we follow the same procedure showed in.[17]

*Ex-situ* visible and Ultra-Violet (UV) Raman characterization was performed in a **z** backscattering geometry by using a Renishaw Invia spectrometer equipped with the 1.96 eV/633nm, 2.41eV/514nm, 2.56eV/488nm and 3.41eV/364nm line of an $Ar^+$ laser line focused on the sample by a 50x 0.75 N.A. Leica objective. The power at the sample was maintained below 1 mW in order to prevent laser induced sample heating and hundreds of spectra have been acquired in order to get the highest signal/noise ratio.

Calculations were carried out in the density-functional theory (DFT) framework, within the generalized gradient approximation (GGA), as implemented in the Quantum ESPRESSO package.[27] The valence electrons of Si were treated using norm-conserving pseudopotentials [28] with a kinetic energy cutoff of 36 Ry. The k-point mesh was set to 4x4x1 during the structural relaxation of the silicene unit cell, which was performed until the average atomic force was lower than $10^{-4}$ Ry/Bohr, while a 25x25x1 Monkhorst-Pack grid was used for the phonon calculations. The grid was then reduced to 3x3x1 for the calculations of the silicene supercells.

The phonon frequency was obtained by diagonalization of the dynamical matrix calculated by the density-functional perturbation theory (DFPT).[29] The non-resonance Raman tensors were obtained within DFPT by second order response to an electric field as implemented in the Quantum ESPRESSO package.[30]

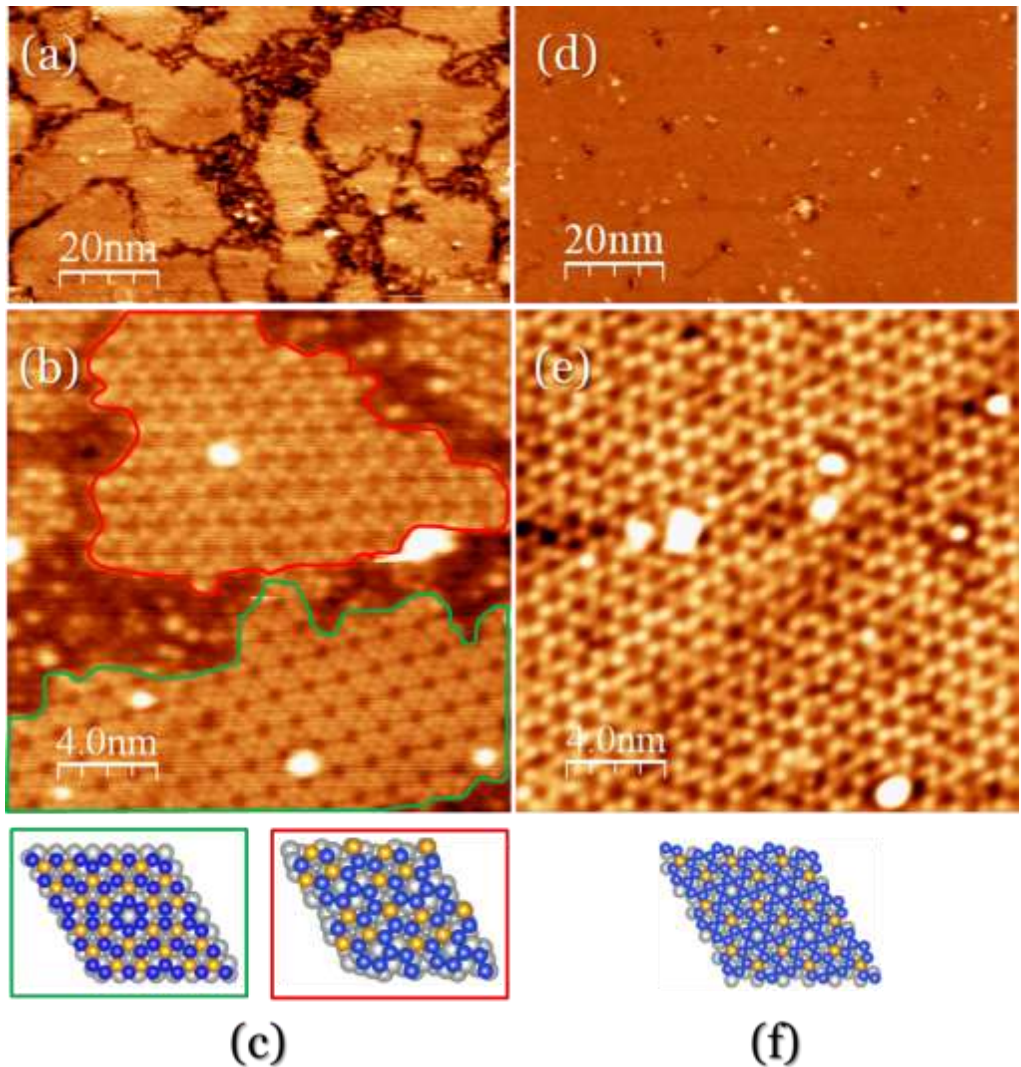

**Figure 1**: (a) Large scale STM characterization of silicene domains on Ag(111) and (b) the two most abundant superstructures: green 4x4, red √13X√13-II;(c) 4x4 and √13X√13-II DFT-GGA relaxed superstructures. (d) Large scale STM characterization and (d) atomic scale topography of the 2√3X2√3 superstructure. (f) DFT-Local Density Approximation relaxed 2√3X2√3 superstructure. STM images were acquired at -1.4 V and 0.4 nA set point. Grey balls are Ag(111) atoms, blue balls are the low lying silicene atoms, yellow balls are the top lying silicene atoms.

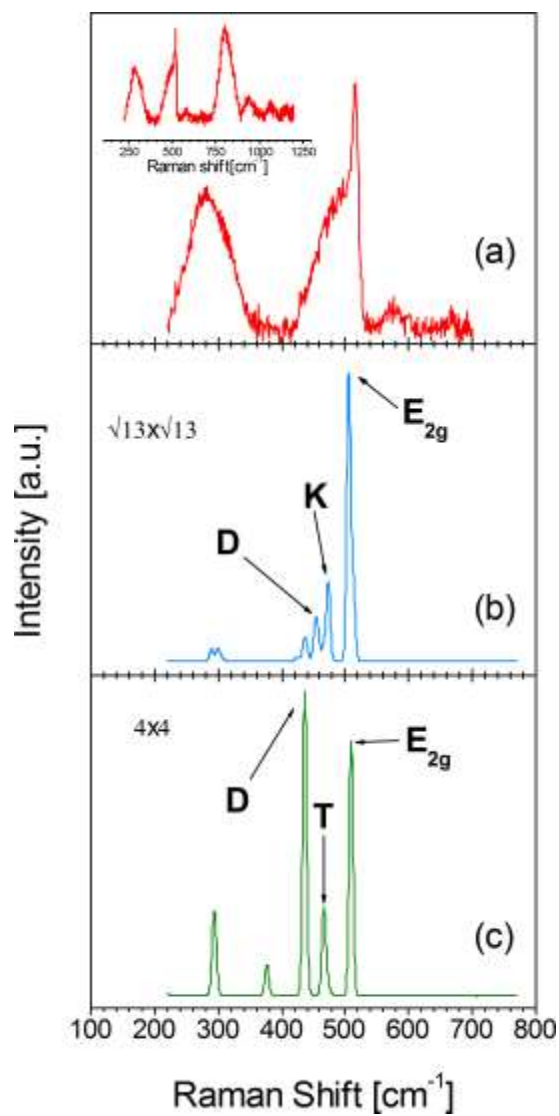

**Figure 2** (a) Experimental Raman spectrum of $Al_2O_3$-capped silicene acquired with the 633 nm laser lines. The inset shows the whole spectrum where at ~300 and ~800 cm$^{-1}$ lie the $Al_2O_3$ features. (b) and (c) computed Raman spectra of the √13X√13-II and 4x4 superstructures respectively, obtained by the calculated vibrational spectra convoluted with a uniform Gaussian broadening having a full width at half maximum (FWHM) of 10 cm$^{-1}$.

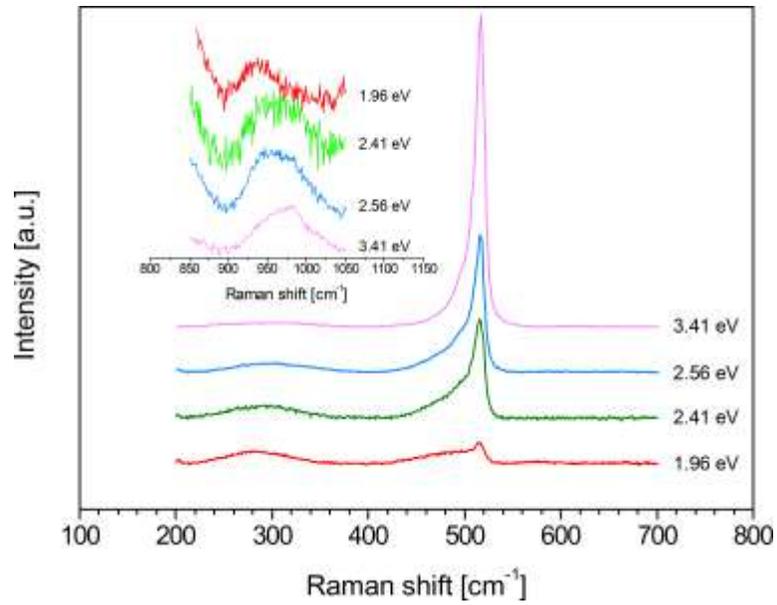

**Figure 3** Raman spectrum of the coupled √13X√13-II and 4x4 superstructures acquired with four excitation energies. The inset shows the dispersion of the band at 900 cm$^{-1}$ as a function of the excitation energy; spectra are vertically stacked for clarity.

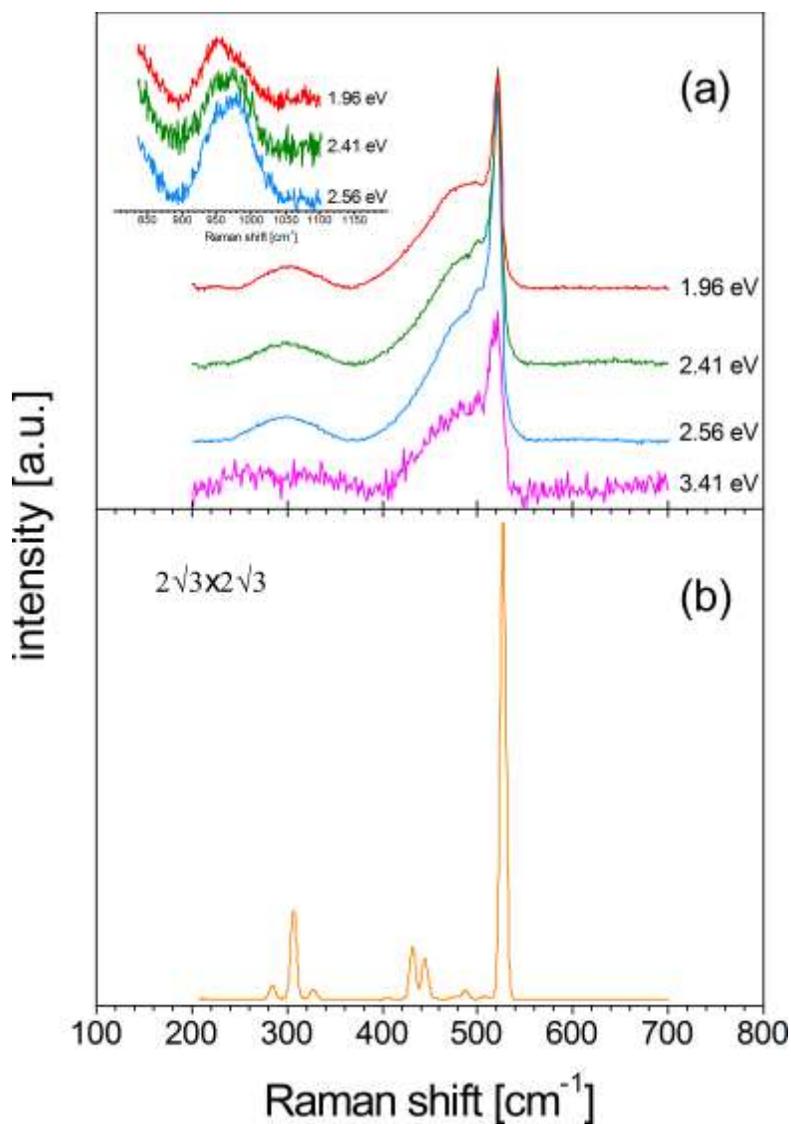

**Figure 4** (a) Raman spectrum of the 2√3X2√3 superstructure acquired with four different excitation energies. The inset shows the dispersion of the band at 900 cm$^{-1}$ as a function of the excitation energy; UV spectrum is not shown because the signal/noise ratio is too low in this spectral region. Spectra are vertically stacked for clarity. (b) computed Raman spectra of the 2√3X2√3 superstructure obtained by the calculated vibrational spectra convoluted with a uniform Gaussian broadening having 10 cm$^{-1}$ FWHM.

The present research activity has been carried on within the framework of the EU project 2D-NANOLATTICES. The project 2D-NANOLATTICE acknowledges the financial support of the Future and Emerging Technologies (FET) programme within the Seventh Framework Programme for Research of the European Commission, under FET-Open grant number: 270749. M.H., E.S. and B. v.d.B. greatly acknowledge Dr. B. Ealet (CNRS and Aix-Marseille University, CINaM) and Dr. G. Pourtois (IMEC and Chemistry Department, PLASMANT group, University of Antwerp).

Supporting Information

Deposition process

Growth experiments have been carried out in an ultra-high vacuum (UHV) system with base pressure in the $10^{-10} mbar$ range incorporating three interconnected chambers for sample processing, chemical analysis and scanning probe diagnostics. The Ag(111) crystal was cleaned by several cycles of $Ar^+$ ion sputtering ($1\ keV$) and subsequent annealing at around 530°C. Si was deposited from a heated crucible (EFM evaporator supplied by Omicron Nanotechnology GmbH) with the substrate at a temperature of 250°C and of 270°. The deposition rate was estimated to be around $6\times10^{-2}$ ML s$^{-1}$. STM topographies were obtained at room temperature using an Omicron STM setup equipped with a chemically etched tungsten tip.

Atom displacements

Figure S1 and S2 reports bond displacements related to Raman active modes of the 4x4 √13X√13-II and 2√3X2√3-II superstructures; next the $E_{2g}$ at 505 cm$^{-1}$(Fig. 4.a), also the disorder-activated $A_{1g}$ modes at 436 cm$^{-1}$, named D (by analogy with the D peak in graphene, see Fig. 4.b) and the T peak at 466 cm$^{-1}$(as it is carried by low standing trimmers, see Fig. 4.c), are activated in the 4x4 silicene superstructures..

On the other hand, the D peak at 455 cm$^{-1}$, together with a $B_{2u}$ at 475 cm$^{-1}$ (named as K in Fig. 4.b), is present also in the calculated spectrum of the √13X√13-II.

The 2√3X2√3-II superstructure presents the $E_{2g}$ mode at 526 cm$^{-1}$ plus three $A_{1g}$ modes at 307 cm$^{-1}$ (planar hexagons), 445 cm$^{-1}$ and 430 cm$^{-1}$ for hexagon with one buckled atom.

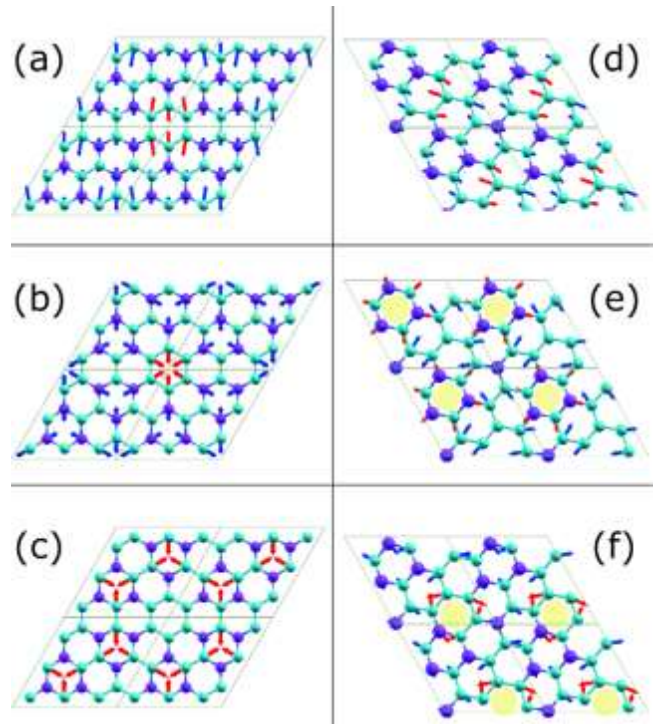

**Figure S1** (a) $E_{2g}$ mode, (b) $A_{1g}$ mode of the planar hexagons and (c) $A_{1g}$ mode of the non-planar hexagons of the 4x4 superstructure. (d) $E_{2g}$ mode, (e) $A_{1g}$ mode and (f) $B_{2u}$ mode of the √13X√13-II superstructure. Light blue atoms are down-standing atoms, dark blue are up-standing atoms. Blue arrow indicates the forces acting on each atom, red arrows indicate forces providing each Raman-active mode. Yellow spots highlight hexagons providing $A_{1g}$ and $B_{2u}$ modes

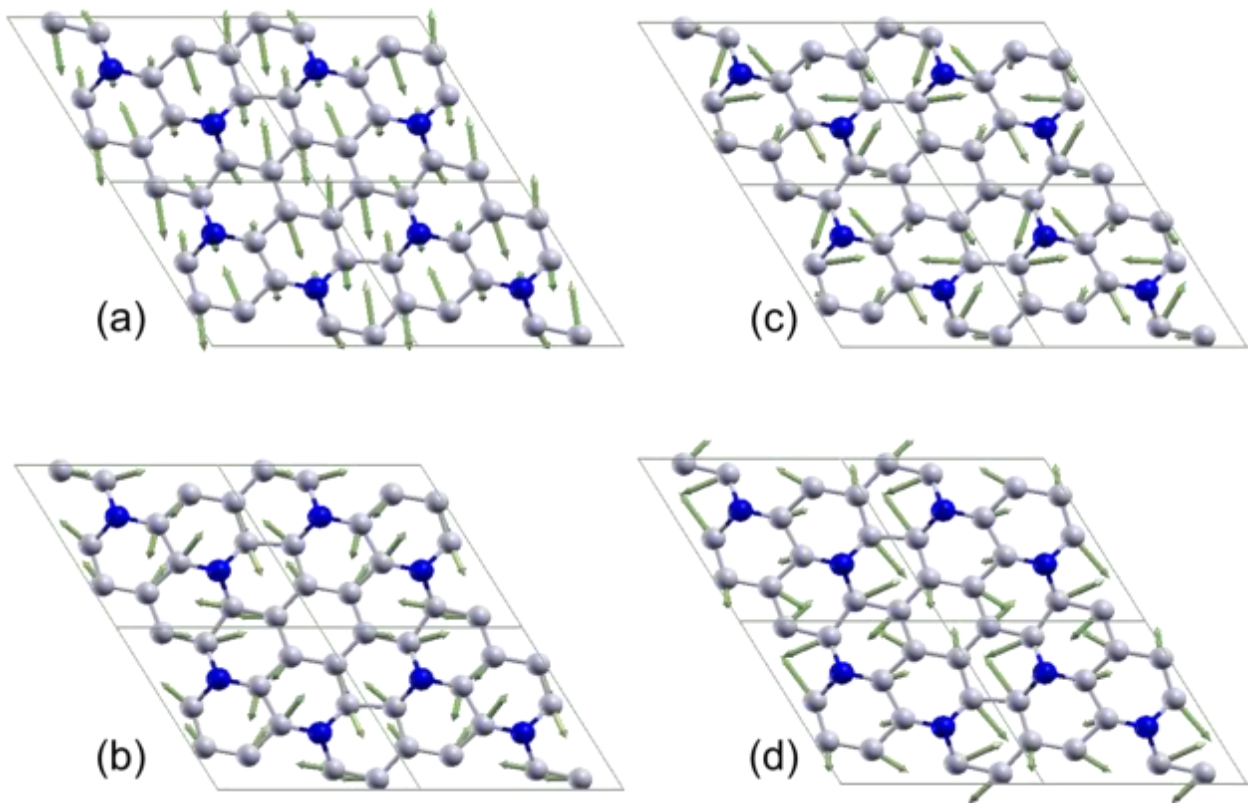

**Figure S2** (a) $E_{2g}$ mode, (b) $A_{1g}$ mode of the planar hexagons, (c) and (d) $A_{1g}$ mode of the non-planar hexagons of the $2\sqrt{3}$X$2\sqrt{3}$-II superstructure. Grey atoms are down-standing atom, blue are up-standing atoms. Green arrow indicates the forces acting on each atom.